\documentclass[reprint,superscriptaddress,groupedaddress,unsortedaddress]{revtex4-1}
\usepackage[pdftex]{graphicx}
\usepackage[english]{babel}
\usepackage{float}
\usepackage{amssymb}   % for math
\usepackage{amsmath}   % for math
\usepackage{epsfig}
\usepackage{dcolumn}   % needed for some tables
\usepackage{bm}        % for math
\newcommand*{\balancecolsandclearpage}{
  \close@column@grid
    \clearpage
      \twocolumngrid
      }

\begin{document}

\title{Toward an improved control of the fixed-node error in quantum Monte Carlo: The case of the 
water molecule}

\author{Michel Caffarel}
\affiliation{Lab. Chimie et Physique Quantiques, CNRS-Universit\'e de Toulouse, France}
\author{Thomas Applencourt}
\affiliation{Lab. Chimie et Physique Quantiques, CNRS-Universit\'e de Toulouse, France}
\author{Emmanuel Giner}
\affiliation{Dipartimento di Scienze Chimiche e Farmaceutiche, Università degli Studi di Ferrara, Ferrara, Italy}
\author{Anthony Scemama}
\affiliation{Lab. Chimie et Physique Quantiques, CNRS-Universit\'e de Toulouse, France}

\begin{abstract}
All-electron Fixed-node Diffusion Monte Carlo (FN-DMC) calculations 
for the nonrelativistic ground-state 
energy of the water molecule at equilibrium geometry are presented. The determinantal part of 
the trial wavefunction is obtained from a perturbatively selected Configuration Interaction 
calculation (CIPSI method) including up to about 1.4 million of determinants.
Calculations are made using the cc-pCV$n$Z family of basis sets, with $n=2$ to 5. 
%For each basis set the CI expansion is sufficiently large to be considered as a good approximation 
%of the FCI solution. 
In contrast with most QMC works no re-optimization of the determinantal part in presence of 
a Jastrow is performed.
For the largest cc-pCV5Z basis set the lowest upper bound for the ground-state energy 
reported so far of -76.43744(18) is obtained. The fixed-node energy is found to decrease 
regularly as a function of the cardinal number $n$ and the Complete Basis Set limit (CBS) 
associated with {\it exact nodes} is easily extracted. The resulting energy of -76.43894(12) -in 
perfect agreement with the best experimentally derived value- is the most accurate theoretical estimate reported so far.
We emphasize that employing selected CI nodes of increasing quality 
in a given family of basis sets may represent a simple, deterministic, reproducible, and systematic 
way of controlling the fixed-node error in DMC.
\end{abstract}
\maketitle

%\balancecolsandclearpage
%Quantum Monte Carlo (QMC) methods are known to be 
%very accurate approaches for electronic structure theory (see, {\it e.g.}
%\cite{hammond_book_1994,foulkes_rmp_2001}). The only uncontrolled source of error\cite{note1} 
The only uncontrolled source of error\cite{note1} in quantum Monte Carlo (QMC) methods
is the fixed-node approximation introduced to suppress the wild fluctuations 
of the sign of the wavefunction (fermion sign problem). 
%In practice, it is realized 
%by restricting the negative and positive walkers to diffuse into separate regions of configuration 
%space delimited by the nodes of some approximate trial wavefunction $\Psi_T$ 
%[nodes= ($3N$-1)-dimensional regions where $\Psi_T(\bf{R})=0$, $N$ number of electrons and $\bf{R}$
%electron spatial coordinates]. Although the fixed-node error is small (typically, 
Although the fixed-node error is small (typically, 
a few percents of the correlation energy), and many fixed-node QMC calculations of impressive 
accuracy have been realized, the error can still be too large in some applications, particularly
in the important case of the computation of (very) small energy differences.

A major challenge for QMC is thus to set up a strategy of 
construction of trial wavefunctions having ``good'' nodes and, even more importantly, to propose
a {\it systematic} way of improving such nodes.
In practice, a standard strategy consists in introducing trial wavefunctions of the best possible 
quality and then to optimize their parameters in a preliminary Variational Monte Carlo (VMC) step 
through minimization of the variational energy or its variance.\cite{umrigar_prl_2007}
%(for example, using the recent linear method of Umrigar {\it et al.}\cite{umrigar_prl_2007}).
Many functional forms for the trial function $\Psi_T$ 
have been explored in the literature,
%(see {\it e.g.}
%\cite{flad_book_1997,filippi_jcp_1996,anderson_jcp_2010,braida_jcp_2011,fracchia_jctc_2012,bouabca_jcp_2010,bajdich_prl_2006,casula_jcp_2004,lopez_pre_2006}). 
the most popular one being the Jastrow Slater form\cite{schmidt_jcp_1990}
\begin{equation}
\Psi_T = e^J \sum_{i} c_i D_i.
\label{eq1}
\end{equation}
combining a Jastrow prefactor $e^J$ containing explicit electronic 
correlations and a short multi-determinantal 
expansion (typically, a few thousands of determinants) describing the multireference 
character of the wavefunction (static correlation effects). 

Very recently some of us have proposed to keep the standard Jastrow Slater form for 
the trial wavefunction but to rely on the more conventional Configuration Interaction (CI) 
expansions of quantum chemistry for the multideterminantal part.
No stochastic re-optimization of the CI expansion is performed, so that ``pure CI'' nodes are employed.
\cite{giner_cjc_1993,scemama_jcp_2014,giner_jcp_2015} The rationale behind this proposal is 
to search for a better control of the fixed-node error by exploiting the unique properties 
of CI wavefunctions. Indeed, CI approaches provide a simple, deterministic, and systematic way 
to build wavefunctions of controllable quality. In a given one-particle basis set,
the wavefunction is improved by increasing the number of determinants, up to the Full 
CI (FCI) limit. Then, by increasing the basis set, the wavefunction can be further improved,
up to the complete basis set (CBS) limit where the exact solution of the electronic Schr{\"o}dinger 
equation is reached. CI nodes, defined as the zeroes of the CI expansions, 
are also expected to display such a systematic improvement.
The main difficulty is of course the exponential 
growth of the space of determinants with respect to the number of electrons and orbitals. 
However, this severe exponential increase can be dramatically attenuated by considering 
{\it Selected} CI (SCI) approaches designed to keep only the most important 
determinants. In practice, we have proposed to 
make use of the CIPSI method (Configuration Interaction 
using a Perturbative Selection done Iteratively),\cite{cipsi1,cipsi2} 
one of the numerous variants of SCI proposed 
in the literature (see, {\it e.g.}, \cite{bender,cipsi1,buenker1,buenker2,buenker3,bruna,buenker-book,cipsi2,harrison}). In this approach the multideterminant 
expansion is built iteratively by selecting determinants according to 
the importance of their second-order perturbational contribution 
to the total energy. As illustrated by a number of applications, CIPSI 
represents a very efficient way of approaching the FCI limit using only a {\it tiny} 
fraction of the total FCI space (see, for example a recent all-electron FCI-converged 
CIPSI calculation for CuCl$_2$ involving 25 electrons and 36 
active orbitals for a FCI space including
10$^{18}$ determinants,\cite{Caffarel_2014}).
This remarkable result is actually common to all 
variants of SCI approaches, including 
the FCI-QMC approach of Alavi {\it et al.}\cite{booth_jcp_2009,cleland_jcp_2010},
which can be considered as a stochastic version of SCI. In practice, 
the main difficulty in using lengthy multideterminant expansions  
in QMC is the expensive cost of evaluating at each Monte Carlo step the first and 
second derivatives of the trial wavefunction (drift vector and local energy). However, 
efficient algorithms have been proposed to perform such 
calculations.\cite{Nukala_2009,clark_jcp_2011,Weerasinghe_2014}
Here, we shall use our recently introduced algorithm allowing to perform converged DMC 
calculations using multideterminant expansions including up to a few millions 
of determinants for a system like 
the water molecule.\cite{preprint_multidets}

A remarkable property systematically observed so far in our first DMC applications using
large CIPSI expansions\cite{giner_cjc_1993,scemama_jcp_2014,giner_jcp_2015} is that, 
except for a possible 
transient regime at small number of determinants,\cite{note2} the fixed-node 
error associated with CIPSI nodes decreases monotonically, both as a function of 
the number of determinants and of the basis set size, leading to the possibility 
of a control of the fixed-node error. Such a result is known
not to be systematically true for a general CI expansion (see, 
{\it e.g.} \cite{flad_book_1997}). However, its validity here could be attributed to the fact that 
determinants are selected in a hierarchical way (the most important ones first), 
so that the wavefunction quality increases step by step, and so the quality of nodes. 

In this Communication all-electron DMC/CIPSI calculations for the water molecule 
at equilibrium geometry using the cc-pCV$n$Z family of basis sets with $n$ ranging from $2$ to 5 and 
large multideterminant expansions including up to 1 423 377 determinants are presented.
The lowest (upper bound) fixed-node energy reported so far of -76.43744(18) is obtained. 
Performing the Complete Basis set (CBS) limit by extrapolating fixed-node energies 
as a function of the cardinal number $n$ of the basis set a 
value of -76.43894(12) for the total energy associated with exact nodes is obtained, in full agreement 
with the best known estimate of -76.4389.\cite{klopper_mp_2001}
%In contrast with most QMC works, no re-optimization of 
%the determinantal part in the presence of a Jastrow is performed here. The nodes used are those 
%of the CI expansions. We emphasize that employing selected CI nodes of increasing quality 
%in a given family of basis sets
%may represent a simple, deterministic, and systematic way of controlling the fixed-node error 
%in DMC.

{\it CIPSI expansion.} 
The multideterminant CIPSI expansion is built by selecting 
iteratively the most important determinants of the FCI expansion. In short 
(for more details, see \cite{giner_cjc_1993}), 
at iteration $n$ the multideterminant expansion $\Psi_D$ is written as the sum of 
the $N^{(n)}$ previously selected determinants (thus, defining the reference space at this iteration)
\begin{equation}
\Psi_D^{(n)}= \sum_{i=1}^{N^{(n)}} c_i^{(n)} D_i
\end{equation}
with energy $
E_0^{(n)}= \frac{\langle \Psi_D^{(n)}|H|\Psi_D^{(n)} \rangle}
                {\langle \Psi_D^{(n)}|\Psi_D^{(n)} \rangle}.$
Then, one determinant (or a group of determinants) $D_j$ 
not belonging to the reference space and corresponding to the greatest second-order 
energy change (or close to it within some threshold),
\begin{equation}
\delta E=-\frac{{\langle \Psi_D^{(n)}|H|D_{j} \rangle}^2}{\langle D_{j}|H|D_{j}\rangle -E_0^{(n)}}
\end{equation}
is (are) selected and added to the reference space. 
At iteration $(n+1)$ the new expansion $\Psi_D^{(n+1)}$ and
energy $E_0^{(n+1)}$ is obtained by diagonalizing 
the Hamiltonian matrix within the new set of selected determinants.
The iterative process is started with the Hartree-Fock determinant or 
a short expansion and is stopped when a target number of determinants is reached. In what follows 
the variational energy associated with the final CI expansion will be denoted as $E_0^{var}$.
%Denoting $|\Psi_D\rangle$ and $E_0^{var}$ the final CI expansion and corresponding 
%variational energy, an estimate of the energy correction 
%to the FCI limit can be obtained by computing the 
%second-order perturbation energy 
%\begin{equation}
%E^{(2)}= - \sum_{j \in M} 
%\frac{{ {\langle \Psi_D|H|D_j} \rangle}^2}{\langle D_j|H|D_j\rangle - E_0^{var}}
%\end{equation}
%where $M$ denotes the set of all determinants not belonging to the reference space
%and connected to the reference vector $|\Psi_D\rangle$
%by $H$ (single and double excitations). The smallest $E^{(2)}$, the better the convergence of the CI 
%expansion is. Finally, a more accurate but non-variational estimate of the FCI energy can be 
%obtained by adding to the variational energy, $E_0^{var}$, the second-order contribution,
%$E_0^{CIPSI} =  E_0^{var} + E^{(2)}.$

{\it Water molecule.} 
In this study we present benchmark calculations for the non-relativistic ground-state
energy of the water molecule at equilibrium geometry, $R_{OH}=0.9572 \AA$ and
$\theta_{OH}=104.52^{o}$. 

{\it CIPSI results.}
All configuration interaction calculations have been carried out using 
our perturbatively selected CI program QUANTUM PACKAGE (downloadable at \cite{quantum_package}).
Standard Dunning type correlation-consistent polarized core-valence basis sets 
cc-pCV$n$Z with $n$ going from 2 to 5 are employed. CIPSI calculations have been performed 
using natural orbitals issued from the diagonalization of the 
one-body density matrix obtained in a preliminary CIPSI run. For each basis set, the selected CI 
expansion has been stopped for one million determinants, except for the largest
cc-pCV5Z basis sets for which two million determinants were considered.
Results are presented in Table \ref{tab1} and compared to the 
recent benchmark CI calculations of Almora-D\`{\i}az including up 
to sextuple excitations.\cite{almora_jcp_2014} 
As we shall see below, truncated versions of these one- and two million-determinant CIPSI expansions 
will actually be used in DMC, results are thus presented for these shorter expansions.
%The values reported in this
%work are expected to approach the FCI limit with an accuracy of at least 
%0.3 mhartree (see, Table III in \cite{almora_jcp_2014})
A remarkable point is the high efficiency of CIPSI in obtaining accurate CI expansion 
with a small number of determinants. For the cc-pCVDZ basis set, the variational energy obtained 
with the 172 256 determinants used in DMC is different from the FCI value of
Almora-D\`{\i}az by only 0.7 mhartree. 
%In view of the estimated error for the benchmark value, 
%both results should be considered as of comparable quality. 
For the other basis sets, the differences remain small, that is 1.8, 1.8, and 2.5 mhartree for the 
cc-pCVTZ, cc-pCVQZ, and cc-pCV5Z basis sets, respectively. 
%We emphasize that for the three basis sets the benchmark values of Almora-D\`{i}az 
%have been obtained with a much greater number of determinants (about $1.6 10^{8}$, 
%$2.3 10^{8}$, and $2.6 10^{8}$, respectively). 
%We emphasize that our goal here is not to get 
%FCI energies of ultimate accuracy but to build CI expansions of the best possible quality 
%with a number of determinants that can be handled in QMC (that is, a maximum of a few millions).
%As it should be, when adding the second-order energy correction, the total energies 
%improve considerably and closely approach the estimated FCI values. 
%(x1,x2,and x3 for the three larger basis sets, respectively). 
%For the largest cc-pCV5Z basis set we were not able to 
%compute $E^{(2)}$ because of computational limits. However, a rough estimate can be obtained 
%by extrapolating the values of the second-order correction computed for increasing numbers of selected determinants to infinite number of determinants.
%A value of about -0.0045 is obtained,
%leading to an energy of -76.4316 in good agreement with the value of -76.4311 reported by Almora-D\`{i}az.
\begin{table*}[]
\begin{center}
\begin{tabular}{llcccccc}
\hline
\hline
Basis set&FCI size&\# dets used in DMC&$E_0^{var}$
&FCI, Almora-D\`{\i}az\cite{almora_jcp_2014} & Deviation \\ 
\hline 
cc-pCVDZ $\;\;\;\;\;$       &$\sim 10^{10}\;\;\;\;\;\;$  & $\;\;$172 256      & 
-76.282136 &
-76.282865 & 0.0007\\
cc-pCVTZ $\;\;\;\;\;$       &$\sim 2.10^{14}\;\;\;\;\;\;$& $\;\;$640 426      & 
-76.388287 & 
-76.390158 & 0.0018\\
cc-pCVQZ $\;\;\;\;\;$       &$\sim 2.10^{17}\;\;\;\;\;\;$& $\;\;$666 927       & 
-76.419324 &
-76.421148 & 0.0018\\
cc-pCV5Z $\;\;\;\;\;$       &$\sim 7.10^{19}\;\;\;\;\;\;$& 1 423 377& 
-76.428550&
-76.431105 & 0.0025\\
\hline
\hline
\end{tabular}
\end{center}
\caption{Number of determinants and corresponding variational energies for CIPSI expansions used 
in DMC for each cc-pCVnZ (n=2 to 5) basis set. Last column: Deviations of the variational energy 
to the best FCI estimates of Almora-D\`{\i}az\cite{almora_jcp_2014}.
Energies in atomic units.}
\label{tab1}
\end{table*}

{\it FN-DMC results.}
All-electron DMC calculations have been realized using our general-purpose QMC program QMC=CHEM
(downloadable at \cite{qmcchem}). A minimal Jastrow prefactor taking care 
of the electron-electron cusp condition is employed and molecular orbitals are slightly modified 
at very short electron-nucleus distances to impose exact electron-nucleus cusp conditions. 
The time-step used, $\tau=2.10^{-4}$ a.u, has been chosen small enough to make
the finite time-step error not observable with statistical fluctuations.
%In view of the very large number of determinants considered here, special attention is required 
%to allow DMC calculations to be feasible in practice. A number of approaches to handle efficiently 
%large multideterminant expansions have been proposed.\cite{Nukala_2009,Clark_2011,Weerasinghe_2014} 
%Here, we employ our recently proposed algorithm presented in detail in \cite{preprint_multidets}. 

To accelerate DMC calculations and not to use the full one- and two million- determinant expansion 
of the initial CIPSI calculations we have employed the improved truncation scheme 
described in \cite{preprint_multidets}.
%In short, the approach consists in removing from the CI expansion written as
%$$ \sum_{i=1}^{N_{\rm det}^\uparrow} \sum_{j=1}^{N_{\rm det}^\downarrow} C_{ij} 
%D^\uparrow_i({\bf R}_\uparrow) D^\downarrow_j({\bf R}_\downarrow)  
%$$
%all determinants $D_K({\bf R})=D^\uparrow_i({\bf R}_\uparrow) D^\downarrow_j({\bf R}_\downarrow)$ 
%containing $\sigma$-determinants ($\sigma=\uparrow,\downarrow$) whose contribution to the norm, 
%$\sum_{j=1}^{{N}_{\rm det}^\downarrow} C^2_{ij}$ for $\uparrow$ electrons 
%or $\sum_{i=1}^{{N}_{\rm det}^\uparrow} C^2_{ij}$ for $\downarrow$ electrons
%is smaller than a given threshold $\epsilon_{\sigma}$. 
In short, the approach consists in writing the CI expansion as
$$ \sum_{i=1}^{N_{\rm det}^\uparrow} \sum_{j=1}^{N_{\rm det}^\downarrow} C_{ij} 
D^\uparrow_i({\bf R}_\uparrow) D^\downarrow_j({\bf R}_\downarrow).
$$
For each $\sigma$-determinant $D^{\sigma}_k$
($\sigma=\uparrow,\downarrow$), the contribution 
to the norm of the wavefunction is given either by 
$\sum_{l=1}^{{N}_{\rm det}^{\downarrow}} C^2_{kl}$ for $\sigma=\uparrow$
or $\sum_{l=1}^{{N}_{\rm det}^{\uparrow}} C^2_{lk}$ for $\sigma=\downarrow$.
If this contribution is
below a given threshold $\epsilon_{\sigma}$ the $\sigma$-determinant
$D^{\sigma}_k$ is discarded ($C_{kl}$ or $C_{lk} = 0\, \forall l$, for the $\uparrow$- or $\downarrow$-sector, respectively). 
Here, $N_{\rm det}^\sigma$ denotes the number of {\it different} 
$\sigma$-determinants in the CI expansion. Such numbers being usually much smaller 
than the total number of products of determinants $D({\bf R})=D^\uparrow D^\downarrow$,
the gain in computational cost can be important (see, 
Table 5 of \cite{preprint_multidets}).
Here, we chose to truncate the expansion 
by taking $\epsilon_{\uparrow}=\epsilon_{\downarrow}=10^{-8}$, except 
for the cc-pCV5Z basis where a value of $10^{-9}$ has been used.
Values for $\epsilon_{\sigma}$ have been 
chosen small enough to get converged fixed-node energies as a function of 
the number of selected determinants within statistical errors. 
In other words, nodes employed in this work are expected to be close to FCI nodes.
The final numbers of selected
determinants used are given in Table \ref{tab1}.

%The actual number of determinants used in DMC and the corresponding CI
%variational energies are given in Table \ref{tab1}. 
\begin{table}[]
\begin{center}
\begin{tabular}{lcc}
\hline
\hline
Basis set[Ndets] &T$_{\rm CPU}$(Ndets)/T$_{\rm CPU}$(1det)& $E_0^{DMC}$\\
\hline 
cc-pCVDZ[172 256]  &$\sim$101. & -76.41571(20)\\
cc-pCVTZ[640 426]  &$\sim$185. & -76.43182(19)\\
cc-pCVQZ[666 927]  &$\sim$128. & -76.43622(14)\\
cc-pCV5Z[1 423 377]&$\sim$235. & -76.43744(18)\\
\hline
\hline
\end{tabular}
\end{center}
\caption{All-electron DMC energies (in a.u.) obtained with CIPSI nodes for each basis set.
Second column: Increase of CPU time due to the use 
of the large multideterminant expansion. 
}
\label{tab2}
\end{table}

\begin{figure}
\begin{center}
\includegraphics[width=\columnwidth]{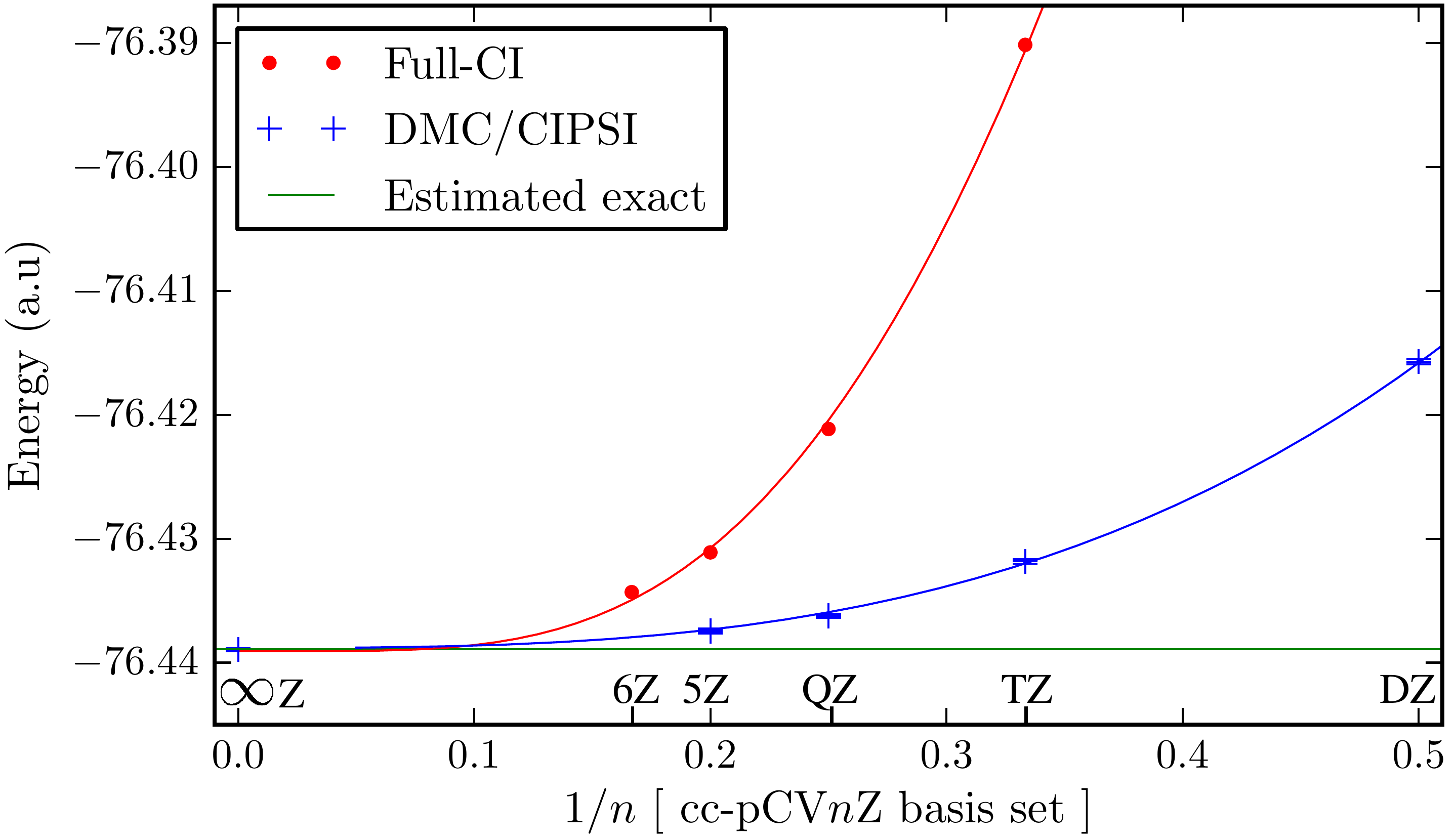}
\caption{CBS extrapolation of FCI and DMC/CIPSI energies. Error bars on DMC data are plotted 
but almost imperceptible.}
\label{figure}
\end{center}
\end{figure}

The efficiency of our algorithm for computing large multiderminant expansions 
can be quantified by measuring the ratio of CPU times 
needed to realize one Monte Carlo step using either the full expansion 
or only the single HF determinant. Such ratios are presented in table \ref{tab2} 
for each basis set.
Fixed-Node DMC energies (in atomic units) obtained with CIPSI nodes for the various basis sets 
are also given in table \ref{tab2}
and plotted in Fig. \ref{figure} as a function of the inverse of the cardinal number $n=2$ to 5. 
The horizontal line is the best estimate of the total nonrelativistic energy reported in the 
literature, see \cite{klopper_mp_2001}. For comparison, we have also reported the best estimates of 
the FCI energies of Almora-D\`{\i}az. Quite remarkably both sets of points display a very similar 
overall behavior. In particular, the values converge smoothly 
to the same CBS limit as a function of the cardinal number $n$ with a typical inverse third 
power law.
Using a simple two-parameter fitting function, $E_0(n)= E_0({\rm CBS}) + a n^{-3}$,
the CBS limit for DMC results gives an extrapolated value of $-76.43894(12)$.
The error bar has been estimated by reproducing the fit over a large statistical ensemble 
of independent data drawn according to their respective error bars. No correlation 
between data being considered, the error value shoud be considered 
as rather conservative. Note that the energy of $-76.43744 \pm 0.00018$ obtained 
with the cc-pCV5Z nodes is the lowest upper bound reported so far in DMC or any other approach.
Regarding computational aspects, calculation of each FN-DMC energies of table \ref{tab2}
were performed using 800 cores on the Curie machine (TGCC/CEA/Genci) during about 15 hours. 
The cost of deterministic CIPSI calculations to build the trial wavefunctions is marginal. Roughly 
speaking, the cost is similar to that needed for making CISD calculations with the same basis sets.

\begin{table}[]
\begin{tabular}{ll}
\hline
\hline
Clark {\it et al.},\cite{clark_jcp_2011} DMC (upper bound)    &      -76.4368(4)    \\
This work, DMC (upper bound) &  -76.43744(18)\\
Almora-D\`{\i}az,\cite{almora_jcp_2014} CISDTQQnSx (upper bound) & -76.4343\\
Helgaker  {\it et al.},\cite{helgaker_jcp_1997} R12-CCSD(T)  & -76.439(2)\\
Muller and Kutzelnigg ,\cite{muller_mp_1997}  R12-CCSD(T)  & -76.4373\\
Almora-D\`{\i}az,\cite{almora_jcp_2014} FCI + CBS  & -76.4386(9)\\
Halkier {\it et al.},\cite{halkier_cpl_1998} CCSD(T)+CBS  & -76.4386\\
Bytautas and Ruedenberg,\cite{bytautas_jcp_2006} FCI+CBS & -76.4390(4)\\
This work, DMC + CBS  &  -76.43894(12)\\
Experimentally derived estimate\cite{klopper_mp_2001} & -76.4389\\
\hline
\hline
\end{tabular}
\caption{\label{tab3} Comparison of nonrelativistic ground-state
total energies of water obtained with the most accurate theoretical 
methods. Energies in a.u.}
\end{table}

In Table \ref{tab3} a selection of the best (lowest) values 
reported in the literature for the total energy of the water molecule is presented.
Using DMC, the lowest value published so far is that of Clark {\it et al.} of
$-76.4368(4)$.  Here, using the nodes of the CIPSI/cc-pCV5Z expansion 
an improved value of $-76.43744(18)$ is obtained. The  
lowest upper bound reached using a post-Hartree Fock correlated approach is that 
of Almora-D\`{\i}az of $-76.4343$, a value significantly higher than DMC values. 
Finally, the best (non-variational) estimates are those obtained by performing CBS extrapolation. 
At FCI level the most accurate one is that of 
Bytautas and Ruedenberg,\cite{bytautas_jcp_2006} $E_0=-76.4390(4)$. Here, our value of -76.43894(12)
is, to the best of our knowledge, the most accurate value reported so far.
In both cases the best experimentally derived 
estimate of -76.4389 is recovered within error bars.

{\it Conclusion}. In this study we have performed DMC calculations 
using nodes of multideterminant CI expansions obtained through a perturbative selection 
of the most important determinants (selected CI). In contrast with most QMC works, no-reoptimization 
of nodes in presence of a Jastrow prefactor has been performed. For each basis set of the 
cc-pCV$n$Z family ($n=2-5$), CIPSI nodes obtained are of near-Full-CI quality.
As a result of the deterministic construction of nodes using CI expansions, 
the total fixed-node energy is found to be a smoothly-decreasing function of the cardinal number 
of the basis set with a typical inverse third power law. The Complete Basis Set (CBS) limit 
leading to the total energy associated with {\it exact nodes} is then easy to perform.
From a general perspective, we emphasize 
that employing selected CI nodes of increasing quality 
in a given family of basis sets may represent a simple, deterministic, reproducible, and 
systematic way of controlling the fixed-node error in DMC.

{\it Acknowledgments.}
AS and MC thank the Agence Nationale pour la Recherche (ANR) for support through 
Grant No ANR 2011 BS08 004 01. This work has been made through generous 
computational support from CALMIP
(Toulouse) under the allocation 2015-0510, and GENCI under the allocation x2015081738.


\begin{thebibliography}{32}
\bibitem [{not({\natexlab{a}})}]{note1}%
  \BibitemOpen
  \href@noop {} {} {\natexlab{a}} \bibinfo {note} {Here, by uncontrolled it
  is meant that there does not exist a systematic way of arbitrarily reducing
  the fixed-node error.}\BibitemShut {Stop}%
\bibitem [{not({\natexlab{b}})}]{note2}%
  \BibitemOpen
  \href@noop {} {} {\natexlab{b}}\bibinfo {note} {An increase of the
  fixed-node energy may be sometimes observed at small number of determinants,
  (say, less than a few thousands) when large basis sets and/or canonical
  orbitals are used. This transient behavior has been found to systematically
  disappear when natural orbitals are used and/or larger expansion are
  considered.}\BibitemShut {Stop}%
\end{thebibliography}
\end{document}